\documentclass{aa}
\usepackage{amsmath}
\usepackage[varg]{txfonts}
\usepackage{graphicx}
\usepackage{epsfig} 
\usepackage{parskip}
\usepackage{wrapfig}
\usepackage{natbib}
\usepackage{rotating}
\usepackage{footnote}
\bibpunct{(}{)}{;}{a}{}{,} 
\usepackage[english]{babel}

\begin{document}

\title{Earth Occultation Imaging of the \\ Low Energy Gamma-Ray Sky with GBM}
\author{J. Rodi\textsuperscript{1} \and M. L. Cherry\textsuperscript{1} \and G. L. Case\textsuperscript{1,2} \and A. Camero-Arranz\textsuperscript{3} \and V. Chaplin\textsuperscript{5} \and \\ M. H. Finger\textsuperscript{4} \and P. Jenke\textsuperscript{5}  \and C. A. Wilson-Hodge\textsuperscript{6}}
\institute{Dept. of Physics \& Astronomy, Louisiana State University, Baton Rouge, LA 70803, USA \and Physics Dept., La Sierrra University, Riverside, CA 92515, USA \and Institut de Ci\`{e}ncies de l'Espai, (IEEC-CSIC), Campus UAB, Fac. de Ci\`{e}ncies, Torre C5, parell, 2a pl., 08193 Barcelona, Spain \and Universities Space Research Association, Huntsville, AL 35805, USA \and Dept of Physics, University of Alabama in Huntsville, Huntsville, AL 35899, USA \and Marshall Space Flight Center, Huntsville, AL 35182, USA}

\abstract {} {The Earth Occultation Technique (EOT) has been applied to \emph{Fermi}'s Gamma-ray Burst Monitor (GBM) to perform all-sky monitoring for a predetermined catalog of hard X-ray/soft \( \gamma\)-ray sources.  In order to search for sources not in the catalog, thus completing the catalog and reducing a source of systematic error in EOT, an imaging method has been developed -- Imaging with a Differential filter using the Earth Occultation Method (IDEOM).} {IDEOM is a tomographic imaging method that takes advantage of the orbital precession of the \emph{Fermi} satellite.  Using IDEOM, all-sky reconstructions have been generated for \( \sim 4 \) years of GBM data in the 12-50 keV, 50-100 keV and 100-300 keV energy bands in search of sources otherwise unmodeled by the GBM occultation analysis.} {IDEOM analysis resulted in the detection of 57 sources in the 12-50 keV energy band, 23 sources in the 50-100 keV energy band, and 7 sources in the 100-300 keV energy band.  Seventeen sources were not present in the original GBM-EOT catalog and have now been added.  We also present the first joined averaged spectra for four persistent sources detected by GBM using EOT and by the Large Area Telescope (LAT) on Fermi: NGC 1275, 3C 273, Cen A, and the Crab.} {}

\maketitle

\section{Introduction}
In 2008 June the \emph{Fermi} satellite was launched with two wide-field-of-view instruments: the Large Area Telescope (LAT) and the Gamma-ray Burst Monitor (GBM), spanning the \( \sim 20 \text{ MeV to } 300 \) GeV \citep{Atwood2009} and 8 keV to 40 MeV \citep{Meegan2009} energy bands, respectively. GBM is currently the only instrument performing all-sky observations above \( \sim 200 \) keV and below the 20 MeV LAT threshold.  The Earth Occultation Technique (EOT) has been applied to the GBM detectors in order to monitor fluxes of known point sources \citep{WilsonHodge2012}.  Over \( \sim 4 \) years of Earth Occultation analysis, the GBM Earth Occultation team has generated daily updated light curves for the \( \sim 200 \) sources currently monitored (available at http://heastro.phys.lsu.edu/gbm/) and published a catalog of high energy (\(>\) 100 keV) sources \citep{Case2011} and a full 3-year catalog of all GBM sources \citep{WilsonHodge2012}.  A particularly surprising result from GBM's continuous all-sky monitoring was the discovery of the variability of the hard X-ray flux from the Crab Nebula \citep{WilsonHodge2011}.

The standard EOT approach requires an input catalog of source positions to monitor sources.  An additional method is necessary for finding sources not in the input catalog.  In order to locate these sources with the uncollimated GBM detectors, a tomographic imaging method has been developed \citep{Rodi2011}.  This technique employs projections of the Earth's limb on the sky over the course of \emph{Fermi}'s orbital precession period (\( \sim 53\) days) to generate all-sky reconstructions.  Imaging with a Differential filter using the Earth Occultation Method (IDEOM) allows for searching for undiscovered sources and known sources absent from the GBM input catalog.  Having a source catalog that is as complete as possible helps reduce systematic errors with EOT and may make it possible to investigate the discrepancies seen at high energies between the Marshall Space Flight Center (MSFC) and the Jet Propulsion Laboratory (JPL) analyses of Burst and Transient Source Experiment (BATSE) data, EOT and the Enhanced BATSE Occultation Package (EBOP) respectively \citep{Harmon2004,Ling2000}.

In this paper we describe the IDEOM imaging method and its application to GBM, and present all-sky reconstructions for the first \( \sim 4 \) years of data in the 12-50 keV, 50-100 keV, and 100-300 keV energy bands.  In Section 2 we give a brief review of the GBM instrument and the standard EOT analysis; Section 3 details IDEOM and its application to GBM; Section 4 presents the imaging results.  Section 5 discusses the four persistent sources detected by GBM and LAT, and in Section 6 we discuss the conclusions and future work.

\section{\emph{Fermi}'s Gamma-ray Burst Monitor}
\emph{Fermi} was launched in 2008 June to an altitude of 565 km and an orbital inclination of \( 25.6^{\circ} \).  At this altitude the Earth's diameter is \( \sim 140^{\circ} \) and thus \( \sim 30 \% \) of the sky is blocked by the Earth at any given time.  Over 85\% of the sky is occulted by the Earth over the course of a single orbit, and the entire sky is occulted in \( \sim 26 \) days (half of the \emph{Fermi} orbital precession period).

GBM consists of 12 NaI detectors and two BGO detectors \citep{Meegan2009}.  The NaI detectors (covering the range 8 keV to 1 MeV) are 12.7 cm in diameter and 1.27 cm thick and are arranged in four groups of three detectors, located on the corners of the spacecraft.  Six detectors are oriented perpendicular to the z-axis of the spacecraft.  Four are inclined \( + 45^{\circ} \) from the +z-direction, and two are inclined \( + 20^{\circ} \) from the +z-direction.  The BGO detectors (covering energies \( \sim 200 \text{ keV to } \sim 40 \) MeV) are 12.7 cm in diameter and 12.7 cm thick, and are located on opposite sides of the spacecraft.  Since 2008, GBM has had two continuous data types, CTIME and CSPEC.  CTIME has nominal 0.256-second resolution and 8-channel spectral resolution.  CSPEC has nominal 4.096-second resolution and 128-channel spectral resolution.

As shown with BATSE on the \emph{Compton Gamma Ray Observatory (CGRO)}, monitoring known hard X-ray/soft gamma-ray sources is possible with simple non-imaging detectors with EOT \citep{Ling2000,Harmon2002}.  As a source is occulted by the Earth, the detector count rate decreases like a step function.  A similar feature, but with opposite polarity, occurs in the detector count rate as the source comes out of occultation.  Occultation times can be predicted using the source coordinates and the spacecraft positions.  The occultation time is defined to be the time at which the transmission through the atmosphere is 50\% at 100 keV.  Given a predetermined catalog of sources, the flux for a source of interest is calculated by selecting a 4-minute window of count-rate data centered on the occultation time and fitted to a quadratic background plus a source transmission model for each source that occults during the window.  A source transmission model is generated that consists of an assumed source spectrum combined with an energy dependent atmospheric transmission model, which is then convolved with the detector response.  For each energy channel for each detector, a scale factor is derived from the fit of the source transmission models to the data and averaged over all of the detectors viewing the source.  The weighted average scale factor for the source of interest is multiplied by the spectral model flux to determine the final flux.  EOT applied to GBM is described in greater detail in \citet{WilsonHodge2012}.  This technique is applied to the daily GBM data producing long-term light curves for all the sources in the GBM input catalog.  In subsequent sections, this process is referred to as the ``daily GBM-EOT analysis.''

\section{Earth Occultation Imaging} 

EOT can monitor only those sources included in the input catalog.  For the \emph{CGRO}/BATSE mission, two tomographic imaging methods were developed, Earth occultation transform imaging \citep{Zhang1993,Zhang1995} and the Likelihood Imaging Method for BATSE Occultation data (LIMBO) \citep{Shaw2004}.  Earth occultation transform imaging generated images with the BATSE data using an inverse Radon transform and a maximum entropy method (MEM) deconvolution.  With this technique, images were created for time periods spanning days to weeks and energies from 20 to 300 keV.  The size of these images ranged from \( 5^{\circ} \times 5^{\circ} \text{ to } 40^{\circ} \times 40^{\circ} \).  Regular analysis consisted of imaging the galactic plane with 27 \( 30^{\circ} \times 30^{\circ} \) fields of view to search for sources.  This method was able to discover a number of transient sources \citep{Zhang1995}, but has characteristics that make systematically searching for sources difficult.  First, the effective image size is limited because of distortions due to approximating the Earth's limb as a straight line instead of an arc, resulting in inaccurate source locations.  A second difficulty is caused by the projections along the limb of bright sources, which can extend for tens of degrees, as discussed below in Sec 3.2.  Earth occultation transform imaging has no way to account for these projections which bias the image background.  Using MEM, which is an iterative, non-linear process, this results in relative source intensities that include spurious contributions of other sources in the field of view, which makes comparing different images difficult.

LIMBO generated all-sky images using a likelihood statistic for a grid of predefined source positions on the sky.  In this method the data were background subtracted using a satellite mass model and then passed through a differential filter before calculating the Maximum Likelihood Ratio for that location.  Sources were then found using a CLEAN deconvolution algorithm.  All-sky images were generated for 489 days of data for a single broad energy band covering 25-160 keV using sky grid points with \( 2^{\circ} \) separation.

The current IDEOM approach makes it possible to locate unmodeled sources by reconstructing the entire sky.  This is important in order to reduce a potential source of systematic errors in the EOT analysis which may arise by attributing flux from an unmodeled source to a source actually in the catalog.  This problem can be seen for example with 1E 1740.7-2942 and GX 1+4 (separated by \( \sim 5.5^{\circ} \)), which occasionally goes into outburst.  If GX 1+4 is absent from the catalog, its transient flux is attributed to nearby sources (e.g. 1E 1740.7-2942).  The second and third columns of Table \ref{complete} show the flux in mCrab of GX 1+4 and 1E 1740.7-2942 from MJD 55715 to 55765 during the time while GX 1+4 is flaring.  The last column shows the flux for 1E 1740.7-2942 during the same period when GX 1+4 is omitted from the catalog.  In the 25-50 keV energy band the calculated flux of 1E 1740.7-2942 more than doubles when GX 1+4 is not included in the catalog, and there is roughly a 35\% increase in flux in the 50-100 keV band.  Thus having a catalog that is as complete as possible is necessary for accurate flux measurements.
\begin{table*}
  \caption{Example of systematic error due to uncatalogued sources}
  \label{complete}
  \centering
  \begin{tabular}{c c c c}
    \hline\hline
      & GX 1+4 & 1E 1740.7-2942 with GX 1+4 & 1E 1740.7-2942 without GX 1+4\\
    Energy Band & Flux (mCrab) & Flux (mCrab) & Flux (mCrab) \\
    \hline
    12-25 keV & \(213.78 \pm  9.35\)  & \(108.97 \pm 15.40\)  & \(111.36 \pm 11.20\) \\
    25-50 keV & \(278.04 \pm 13.07\) &  \( 71.95 \pm 21.05\)  & \(161.13 \pm 16.42\) \\
    50-100 keV &\(185.24 \pm 19.53\) &  \( 95.37 \pm 31.26\)  & \(128.39 \pm 23.99\) \\
    \hline
  \end{tabular}
\end{table*}
    
Another motivation is to investigate a potential cause of the discrepancies between the MSFC (EOT) and JPL (EBOP) analyses of the BATSE occultation data at high energies (100-300 keV).  Results from the JPL method \citep{Ling2000} have shown larger high energy fluxes for some sources than corresponding results from the MSFC analysis \citep{Harmon2004}.  These apparent high energy tails appear correlated with the phase in the 52-day precession period of the satellite, indicating a systematic effect possibly due to unmodeled sources.

\subsection{IDEOM with GBM}

IDEOM builds images of the sky using projections of the Earth's limb on the sky.  Because the detectors are uncollimated, the detector count rate gives a measure of the sky brightness of the visible sky.  Thus for a detector facing the Earth, the count rate consists of a term proportional to the change in the integrated flux along the Earth's limb \citep{Shaw2004} superimposed on a time-varying background and the contribution of sources in the sky moving across the face of each detector.  For a fixed point on the sky observed during rise and set occultation steps on successive orbits, the flux (change in the line integral along the limb) is measured along different angles with respect to the sky.  During the orbital precession period of the satellite, a source's elevation angle \( \beta \) (the angle between the source as it occults and the orbital plane of the satellite (Harmon et al. 2002)) changes thus sampling a range of limb projections for that point on the sky.

\begin{figure*}[t]
\vspace{5mm}
\centering
\includegraphics[angle=90,scale=0.65, trim=0mm 50mm 80mm 0mm]{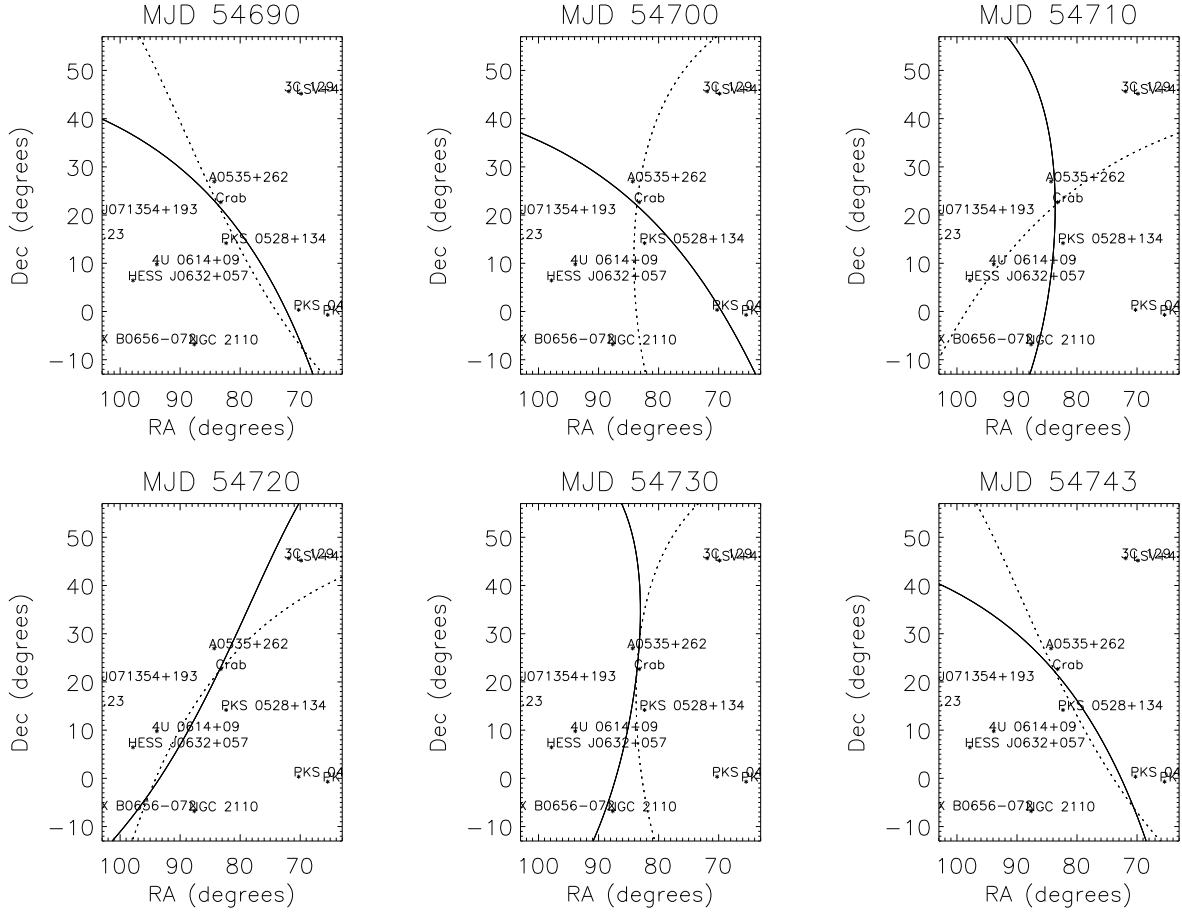}
\vspace{10mm}
\caption{The rise and set limb orientations for the Crab for a single orbit at approximately 10 day intervals during a precession period.  The solid line marks the rise limb, and the dotted line marks the set limb. \label{limbs}}
\end{figure*}
A reconstruction of the sky is then generated by summing the limb projections for a location on the sky during an orbital precession period.  Fig.~\ref{limbs} shows sampled projections for the Crab from a single orbit at \( \sim 10\) day intervals during the precession period from 2008 August 12 (MJD 54690) to 2008 October 4 (MJD 55743).  The solid line marks the rise limb, and the dotted line marks the set limb.  The rise and set limbs cross at the location of the Crab.  Nearby sources from the GBM input catalog are plotted as well.  As the \( \beta\) angle changes, the rise/set limb projections cross the Crab's position at a different orientation on each occultation.

The top row in Fig.~\ref{limbspp} shows sampled projections for two orbits from each day during the same precession period for 3C 273 (\( \delta = 2.05^{\circ}\)), the Crab (\( \delta = 22.01^{\circ}\)), and NGC 4151 (\( \delta = 39.42^{\circ}\)), respectively.  The maximum \( \beta \) angle is declination dependent.  As a result, {more angles are sampled as \( \delta\) increases, as shown in Fig.~\ref{limbspp}.  The bottom row shows the limb projections summed over an entire orbital precession period for a test source of unit intensity corresponding to the location of the source in the top row - i.e., the point spread function for that location on the sky.
\begin{figure*}
\centering
\includegraphics[angle=90,scale=0.65, trim=0mm 50mm 30mm 0mm]{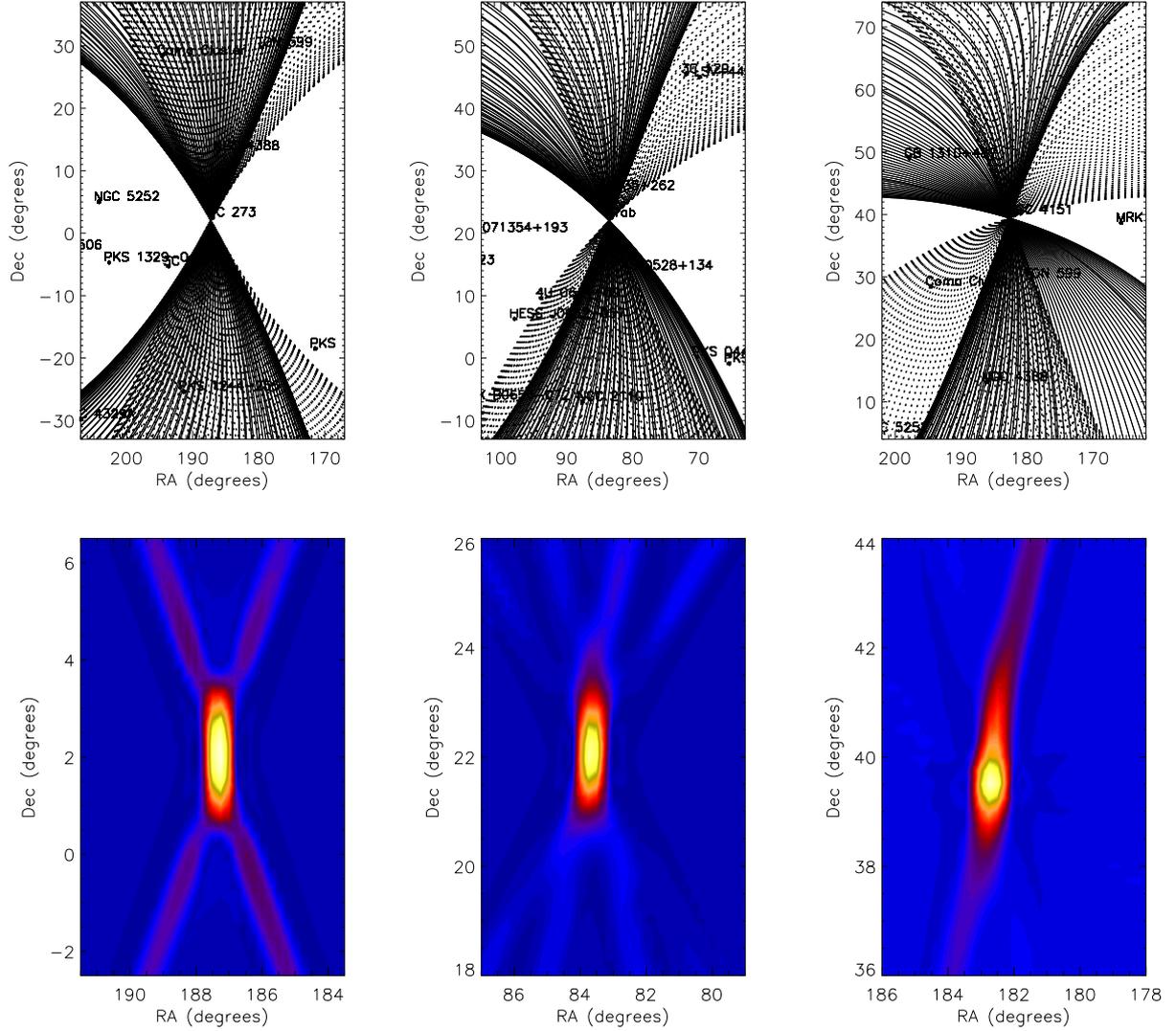}
\vspace{10mm}
\caption{(\emph{Top row}) The rise and set limb orientations for an entire precession period for 3C 273 (left), the Crab (center), and NGC 4151 (right).  (\emph{Bottom row}) Point spread functions for test source at location of actual source. \label{limbspp}}
\end{figure*}

In the energy range corresponding to the GBM NaI detectors (8 keV to 1 MeV), a single step has a finite duration of \( \sim 8 \text{ s } / \cos \beta \) as a source passes behind the Earth's atmosphere \citep{Case2011,WilsonHodge2012,Harmon2002}.  (A derivation of the occultation duration can be found in Appendix B of \citet{Harmon2002}.)  The best possible position resolution depends on the angle subtended by the Earth's atmosphere and can be expressed as \( \Delta \theta = \frac{0.5^{\circ}}{ \cos \beta} \).  \emph{Fermi}'s orbital period of \( \sim 96\) minutes corresponds to a minimum \( \Delta \theta \sim 0.5^{\circ} \).  \( \Delta \theta \) increases as \( \beta \) increases until \( \beta \gtrsim 66^{\circ} \), at which point sources are no longer occulted.

IDEOM creates an input catalog of \( \sim 600,000\) virtual sources uniformly covering the entire sky with \( 0.25^{\circ}\) spacing.  For each virtual source, IDEOM then combines all the projections from a precession period to calculate the intensity at that location.  Energy dependent occultation times are predicted for all the virtual sources so that the occultation time occurs at 50\% transmission in a given energy band.  A 4-minute window of GBM CTIME counts data is selected for the detectors that view each occultation at less than \( 75^{\circ} \) from the detector normal.  A first order detector response correction is applied to account for detector efficiency as a function of the photon energy and the angle between the source and detector normal.  Combining both rise and set occultation step types provides observations at more projection angles.  In order to combine rise steps and set steps, windows with rise steps are converted to set steps by rotating the window about the occultation time at the center of the window.  The weighted average of all windows for each individual virtual source is calculated over an orbital precession period.  Because the occultation windows are in units of counts, Poisson errors are assumed for combining occultation windows where \(\sigma_i = \frac{ \sqrt{N_i}}{R}\), where \( \sigma_i \) is the error in the measurement of the number of counts \(N_i\) in an individual window with response correction \(R\).  Error values are then propagated through the rest of the analysis.  

This average window is rebinned from the standard CTIME 0.256-second resolution to 2.048-second resolution. A differential filter (See Fig.\ref{fig:crab}) as used by \citet{Shaw2004} then subtracts the sum of \( f_b \) bins to the right of the central gap from the sum of \( f_b \) bins to the left with an inner gap of \( 2 f_a \) bins width between the two sums, i.e.
\begin{equation}
o_i = \frac{ \sum_{j=i+f_a}^{j=i+f_a+f_b} r_j - \sum_{j=i-f_a-f_b}^{j=i-f_a} r_j}{f_b}.
\end{equation}
Here \( r_j \) is the number of counts in bin \( j \) and \(o_i\) is the difference between the two sums divided by \(f_b\) for a given bin.  To maximize sensitivity the inner bound should be large enough that \( 2 f_a \) is roughly equal to the occultation duration thus maximizing the difference between total transmission and total attenuation. Decreasing the outer bound \(f_b\) increases the angular resolution but decreases the statistical precision.  For a constant value of \(f_a\), the filtered results are not strongly dependent on the value of \(f_b\).   In practice, we use \( f_a = 3 \text{ and } f_b = 8 \) in order to minimize effects from fluctuations and maximize sensitivity.  

Differential filtering produces a dip at the occultation time while smoothing out background features, as shown in Fig.~\ref{fig:crab} for the Crab.  The source intensity is calculated from the amplitude of the dip by fitting bins within \( \pm (2 f_a + f_b) \) of the occultation time to a polynomial (using the best fit results from either a quadratic or a cubic polynomial), and bins outside this region to a spline function joined by a straight line.  The amplitude of the virtual source is found by taking the difference between the two fits at the occultation time.  A region of \( \pm (2 f_a + f_b) \) is used to ensure that the dip is fully enclosed and thus the difference between the dip and the background is optimally sampled.  The error on the amplitude is taken to be the combined error in the spline and polynomial fits at the peak.  The bottom panel of Fig.~\ref{fig:crab} shows the filtered window with the polynomial fit in (green) X's and the spline fit in (red) triangles.  The outer dashed lines are the boundaries for the polynomial fit, and the middle dashed line denotes the occultation time.  

Cuts are made on the raw data to exclude times when the spacecraft has a high spin rate and when it passes through the South Atlantic Anomaly (SAA).  High spin-rate effects occur when the spacecraft slews, which leads to a rapidly varying background during the 4-minute window.  When passed through the differential filter, a rapidly varying background can mimic an occultation step, creating artifacts in the image.  The SAA passage effect arises from activation in the detectors during passes through the SAA and results in large broad spikes in the data, which again leads to a rapidly varying background.

\begin{figure*}[t]
\centering
\includegraphics[angle=90,scale=0.6, trim=75mm 0mm 40mm 20mm]{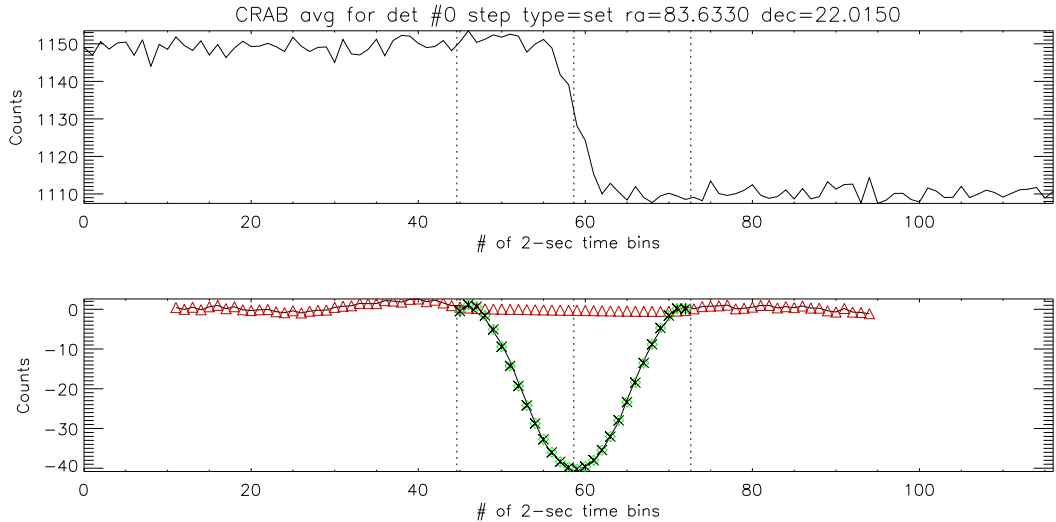}
\vspace{10mm}
\caption{\emph{Top}: Averaged window for 10 days for the Crab.  \emph{Bottom}: The filtered data with the central portion fit to a polynomial (green stars), and the outer background portion fit to a spline function (red triangles). (A color version of this figure is available in the online journal.)\label{fig:crab} }
\end{figure*}

\subsection{Systematic Effects}
Two systematic effects are explicitly taken into account in the IDEOM processing: 1) source confusion with bright sources and 2) windows in which the rise and set steps of a bright source dominate the spline background fit.  The issue of source confusion arises from the ambiguity of where along the Earth's limb the measured flux originates.  Since the GBM detectors have no direct imaging capability, the flux for an occultation time can be attributed to any location on the sky along the Earth's limb.  Consequently, bright sources (e.g. Crab, Cyg X-1, Sco X-1) produce a characteristic ``X'' pattern (bottom row of Fig.~\ref{limbspp}) that can extend for tens of degrees on  the sky.  These ``arms'' are the result of projections remaining in a similar orientation for a relatively long time as the source reaches its maximum and minimum \( \beta \) angles during the \emph{Fermi} precession period, and may overwhelm any flux that comes from possible faint sources lying on or near one of these arms.  To mitigate the effects of source confusion, an algorithm has been developed to ignore occultation windows for virtual sources when a bright source from a pre-determined list occults close in time (\( < 11 \) seconds) to the virtual source but is far away on the sky ( \( \gtrsim 10^{\circ} \)).  

The other systematic effect involves virtual sources at roughly the same declination as a bright source and a few degrees away in right ascension.  The virtual source is close enough on the sky to the bright source so that both rise and set steps for the bright source occur inside the 4-minute window, but the two sources are far enough away that when the window is filtered, the rise steps and the set steps for the bright source remain separated. An example can be seen in Fig.~\ref{hole}  for a virtual source at \( \alpha = 84.75^{\circ} \text {, }   \delta = 21.0^{\circ} \). (The Crab is at \( \alpha = 83.63^{\circ} \text{, } \delta = 22.02^{\circ} \).)  The steps from the bright source are included in the background level, so that the calculated amplitude of the virtual source appears negative.  To reduce this effect and make the calculated amplitude less negative, the filter boundaries are set as usual, but the central region is fitted over a region of only \( \pm f_a \) instead of the normal \( \pm (2 f_a + f_b) \).  This reduces the effect of the bright source on the background fit making the calculated amplitude less negative.  The bottom panel of Fig.~\ref{hole} shows the effect of restricting the central fit region to bins 59-65.

\begin{figure*}[t]
\centering
\includegraphics[angle=90,scale=0.6, trim=0mm 45mm 50mm 0mm]{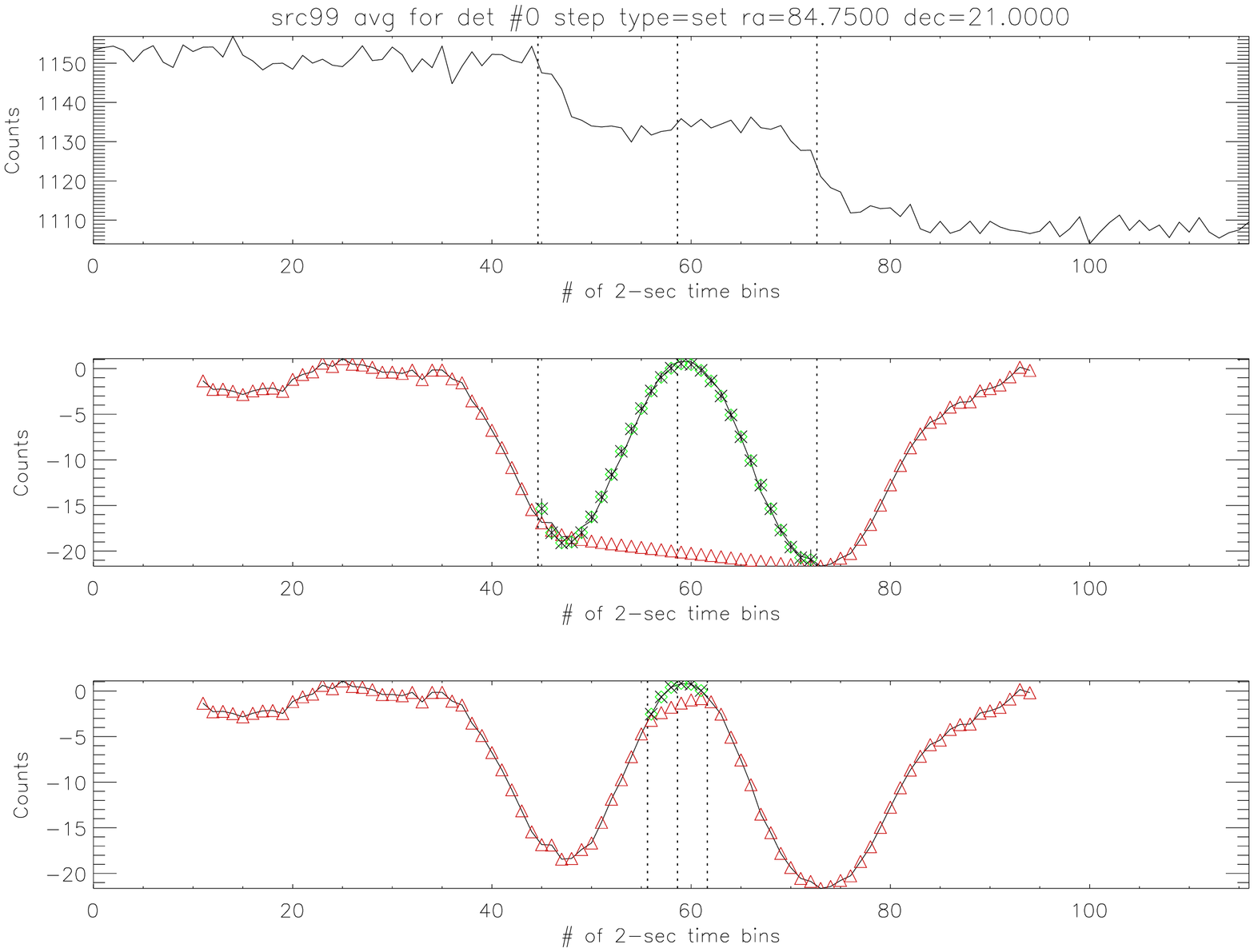}
\vspace{10mm}
\caption{\emph{Top}: Averaged window for 10 days for a virtual source near the Crab.  Crab steps are at bins \( \sim 47 \text{ and } \sim 72\).  \emph{Middle}: The filtered data with the central portion fit to a polynomial (green), and the outer background portion fit to a spline function (red).  Large dips from Crab steps result in negative amplitude for the virtual source.  \emph{Bottom}: Filtered data refit with central portion only \( \pm f_a \) instead of the normal \( \pm (2 f_a + f_b) \).}
\label{hole}
\end{figure*}

\subsection{Differences between GBM-EOT Daily Analysis and IDEOM}
EOT uses an input catalog of positions of known point sources. This restricts EOT to monitoring only sources in its input catalog.  The input catalog for IDEOM consists of virtual sources with \(0.25^{\circ}\) spacing on the sky that may not correspond to the positions of known point sources.  EOT builds a model for the source of interest plus all other bright sources (sources with flux > 50 mCrab in the 12-25 keV band) in the window, and then fits that model to the observed counts in the window.  IDEOM does not fit windows on a step by step basis, instead averaging all the windows over a precession period.  IDEOM is intended to identify unknown sources not included in the original EOT catalog. However, the differences between the EOT and IDEOM methods result in a difference in sensitivity which is due to several factors.  First, EOT uses the exact source position and thus is able to optimally sample the occultation step (and flux) from a source.  The IDEOM virtual source is not necessarily located at the actual source location and so may not sample the occultation step as well.  Also, the interference time is larger for IDEOM (11 s) compared to EOT (8 s) which can result in fewer usable occultation steps for some source locations on the sky (e.g. near Sco X-1 arms).  Finally, in order to reduce potential problems after SAA passes, the data within 400 s after an SAA passage have been ignored in IDEOM likely losing some valid occultation steps. These effects reduce source confusion from distant sources and minimize rapid background changes which may appear as spurious steps in the counting rate, but mean that IDEOM cannot distinguish sources that are close together on the sky as well as EOT.

In some situations, especially for sources with hard spectra, IDEOM may have better sensitivity than the GBM-EOT daily analysis.  The GBM-EOT analysis excludes occultation windows when a source whose 12-25 keV flux is greater than 50 mCrab occults within 8 s of the source of interest.  The EOT sensitivity then is low at all energies for sources close together as the number of usable occultation steps for each source is reduced based on the 12-25 keV flux.  Since IDEOM does not generally exclude occultation steps for sources close together, analysis by IDEOM of independent energy ranges can detect sources at higher energies that may have lost occultation steps in EOT.  An example is the case of GRS 1758-258 (\( \sim 0.75^{\circ} \) away from GX 5-1), described below in the discussion of Table 4.

\section{Results}

\begin{table}
  \caption{\emph{Fermi} Orbital Precession Periods}
  \label{periods}
  \centering
  \begin{tabular}{c c c}
    \hline\hline
    Precession Period & Beginning Time  & Ending Time \\
                      & (MJD)          & (MJD)      \\
    \hline
    1 & 54690.0 & 54744.0  \\
    2 & 54744.0 & 54798.0  \\
    3 & 54798.0 & 54852.0  \\
    4 & 54852.0 & 54906.0  \\
    5 & 54906.0 & 54959.0  \\
    6 & 54959.0 & 55013.0  \\
    7 & 55013.0 & 55067.0  \\
    8 & 55067.0 & 55121.0  \\
    9 & 55121.0 & 55175.0  \\
    10 & 55175.0 & 55228.0  \\
    11 & 55228.0 & 55282.0 \\
    12 & 55282.0 & 55336.0 \\
    13 & 55336.0 & 55390.0 \\
    14 & 55390.0 & 55443.0 \\
    15 & 55443.0 & 55497.0 \\
    16 & 55497.0 & 55551.0 \\
    17 & 55551.0 & 55605.0 \\
    18 & 55605.0 & 55651.0 \\
    19 & 55651.0 & 55705.0 \\
    20 & 55705.0 & 55758.0 \\
    21 & 55758.0 & 55812.0 \\
    22 & 55812.0 & 55865.0 \\
    23 & 55865.0 & 55918.0 \\
    24 & 55918.0 & 55972.0 \\
    25 & 55972.0 & 56025.0 \\
    26 & 56025.0 & 56078.0 \\
    27 & 56078.0 & 56131.0 \\
    28 & 56131.0 & 56185.0 \\
    \hline
\end{tabular}
\end{table}

IDEOM has been applied to the first \( \sim 4 \) years of GBM CTIME data, covering 2008 August 12 to 2012 September 14, which corresponds to 28 \emph{Fermi} orbital precession periods.  An all-sky reconstruction has been generated for each precession period for three broad energy bands.  Table~\ref{periods} shows the beginning and ending dates for each precession period.  For the beginning and ending MJD, the limb projections have been summed over the entire day regardless of the fractional part of the precession period for that day.  Precession period 22 was excluded from the maps of the lowest energy band due to high significance fluctuations in the background.  Also, the high mass X-ray binary A0535+262 was in an extremely bright outburst phase at times during precession periods 9, 10, and 18. Flickering of the source during these times resulted in wave-like features contaminating the reconstructions; these times  were also removed.

Single precession period reconstructions for an energy band were combined by summing the intensities for a virtual source weighted by the number of occultation steps for the point during a precession period.  Figs.\ ~\ref{fig:map12}-\ref{fig:map100} show the combined significance maps for the first \( \sim 4 \) years of the mission for the 12-50 keV, 50-100 keV, and 100-300 keV energy bands, respectively.  Green contours have been over plotted at \( 3.5 \sigma, 5 \sigma, 10 \sigma, \text{ and } 20 \sigma \).  Features often extend from bright sources (e.g. Crab, Vela X-1, GRS 1915+105) which show secondary effects from limb projections adding constructively during a precession period.  These ``arms'' are artifacts of Earth occultation imaging methods.  The ``streams'' near (\(180^{\circ} , 30^{\circ}\)) are likely due to incomplete removal of Crab occultation limbs.

Initially (Method 1), candidate sources were identified in the IDEOM sky maps if a) their peak significance exceeded $3.5 \sigma$ above background, b) they were located within $0.75^\circ$ of a known source in the \emph{Swift}/BAT, \emph{INTEGRAL}/SPI, or \emph{Fermi}/LAT catalog, and c) their statistical significance in the 12-50 keV band was confirmed by EOT at a level of $10 \sigma$.  Seventeen sources were found with this approach and have been added to the GBM EOT catalog.  These sources are plotted in Fig.\ ~\ref{fig:map12} as blue triangles.

In addition (Method 2), a search for potential sources was performed whereby the sky was divided into \( 5^{\circ} \times  5^{\circ} \) regions centered on each point. Independently in the 12 - 50, 50  - 100, and 100 - 300 keV energy bands, the reconstructed intensity in each \( 5^{\circ} \times  5^{\circ} \) region was fit to a two-dimensional Gaussian.  For potential sources with peak significance $> 3.5 \sigma$ over background, the centroid of the Gaussian was taken to be the source position.  The integrated significance was then calculated over a rectangle \( \pm 2 \sigma_{\alpha,\delta} \) around the source position, where \( \sigma_{\alpha} \) and \( \sigma_{\delta} \) are the calculated widths of the fitted Gaussian, by summing the counts value for each virtual source in the region divided by the summed error for all of the virtual sources.  This reduces the weight of single ``hot'' pixels.  Sources with an integrated significance $ > 14 \sigma$ were then added into a temporary source catalog and re-analyzed with EOT; sources that were confirmed with $> 5 \sigma$ total significance (statistical + systematic) by EOT were accepted as real sources.  Sources detected \(> 5 \sigma\) by the daily GBM-EOT analysis are plotted as red asterisks in Figs.\ ~\ref{fig:map12}-\ref{fig:map100}.  Sources detected by Method 2 but not detected by the daily GBM-EOT analysis are plotted as black triangles in Figs.\ ~\ref{fig:map12}-\ref{fig:map100}.  For clarity, Table~\ref{galridge} lists sources detected by the daily GBM-EOT analysis along the Galactic Ridge in the 12-50 keV range; those sources can be identified in Fig.~\ref{fig:map12} by the corresponding number.

\begin{table}
  \caption{Sources Detected by Daily GBM-EOT Analysis along Galactic Ridge}
  \label{galridge}
  \centering
  \begin{tabular}{c c}
    \hline\hline
    Source Number & Source Name \\
     &             \\
    \hline
    1  & X1624-490         \\
    2  & IGR J16318-4848   \\
    3  & AX J1631.9-4752   \\
    4  & 4U 1630-472       \\
    5  & 4U 1636-536       \\
    6  & GX 340+0          \\
    7  & OAO 1657-415      \\
    8  & 4U 1700-377       \\
    9  & GX 349+2          \\
    10 & 4U 1702-429       \\
    11 & H1705-440         \\
    12 & 4U 1708-407       \\
    13 & GX 9+9            \\
    14 & GX 354-0          \\
    15 & GX 1+4            \\
    16 & H1730-333         \\
    17 & KS 1731-260       \\
    18 & SLX 1735-269      \\
    19 & X1735-444         \\
    20 & 1E 1740.7-2942    \\
    21 & MAXI J1745-288    \\
    22 & 1A 1742-294       \\
    23 & IGR J17464-3213   \\
    24 & IGR J17473-2721   \\
    25 & GX 3+1            \\
    26 & 4U 1746-370       \\
    27 & GX 5-1            \\
    28 & GRS 1758-258      \\
    29 & GX 9+1            \\
    30 & SAX J1806.5-2215  \\
    31 & GX 13+1           \\
    32 & GX 17+2           \\
    33 & H1820-303         \\
    34 & GS 1826-238       \\
    \hline
\end{tabular}
\end{table}

\begin{figure*}
\vspace{10mm}
\centering
\includegraphics[angle=180,scale=0.9, trim=0mm 0mm 70mm 30mm]{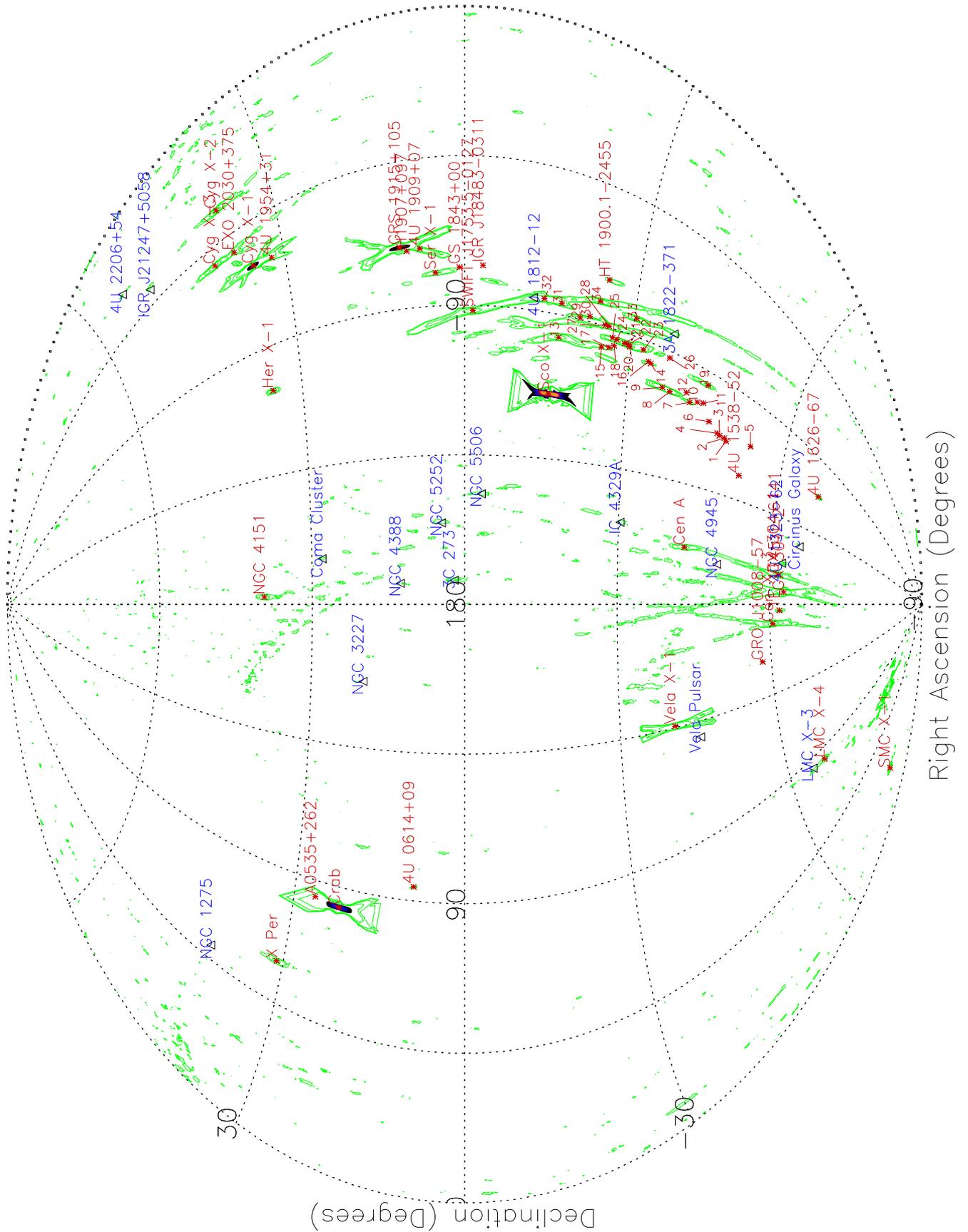}
\vspace{75mm}
\caption{All-sky reconstruction in the 12-50 keV band for 2008 August 12 to 2012 September 14.  Sources marked with black triangles denote sources added to the GBM catalog based on imaging work.  Contours have been over plotted at 3. 5,  5,  10,  and 20 sigma.  Sources labeled with numbers are identified in Table~\ref{galridge}.}
\label{fig:map12}
\end{figure*}
\begin{figure*}
\vspace{10mm}
\centering 
\includegraphics[angle=180,scale=0.9, trim=0mm 0mm 70mm 30mm]{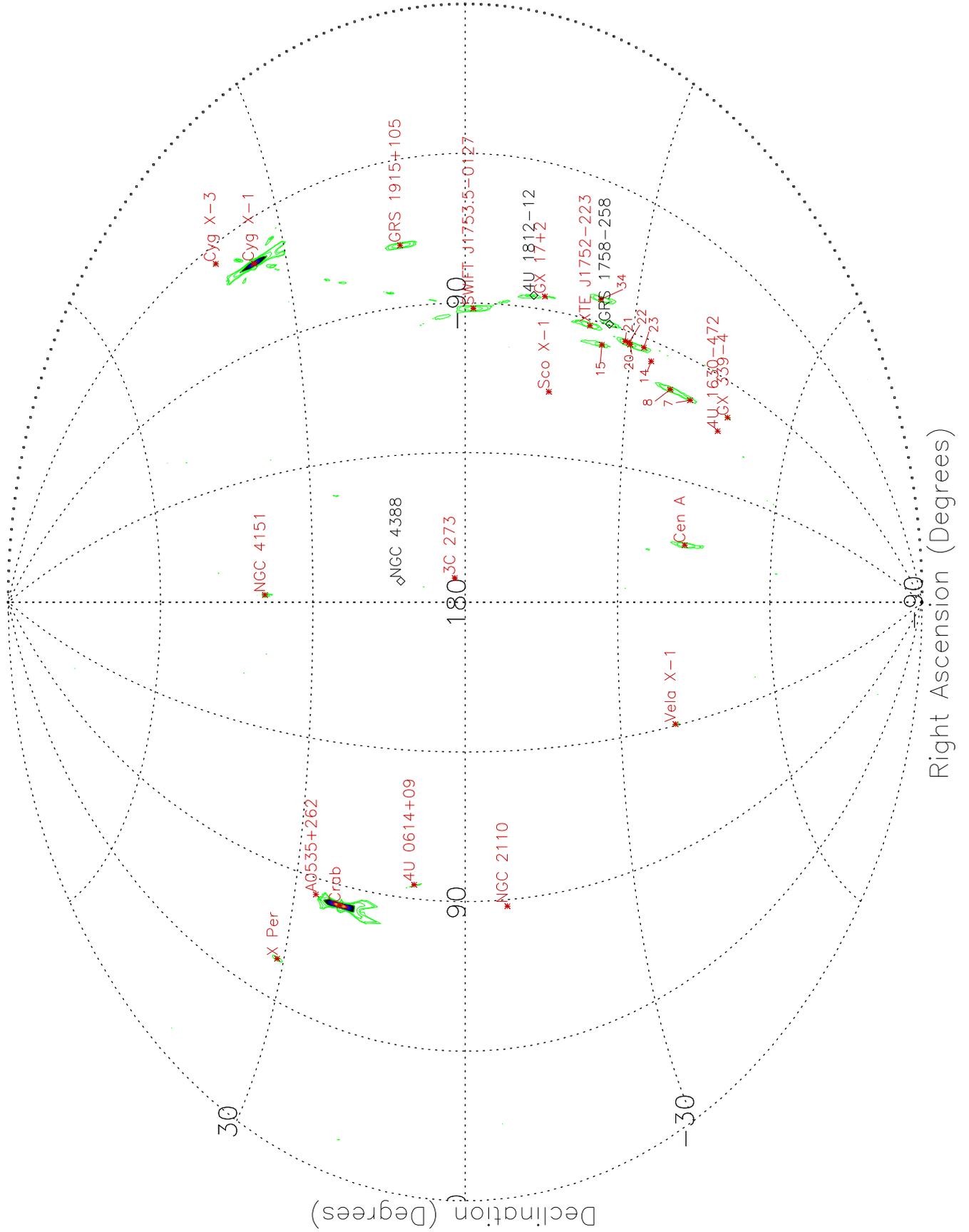}
\vspace{75mm}
\caption{All-sky reconstruction in the 50-100 keV band for 2008 August 12 to 2012 September 14.  Contours have been over plotted at 3. 5,  5,  10,  and 20 sigma.  Sources labelled with numbers are identified in Table~\ref{galridge}.}
\label{fig:map50}
\end{figure*}
\begin{figure*}
\vspace{10mm}
\centering
\includegraphics[angle=180,scale=0.9, trim=0mm 0mm 70mm 30mm]{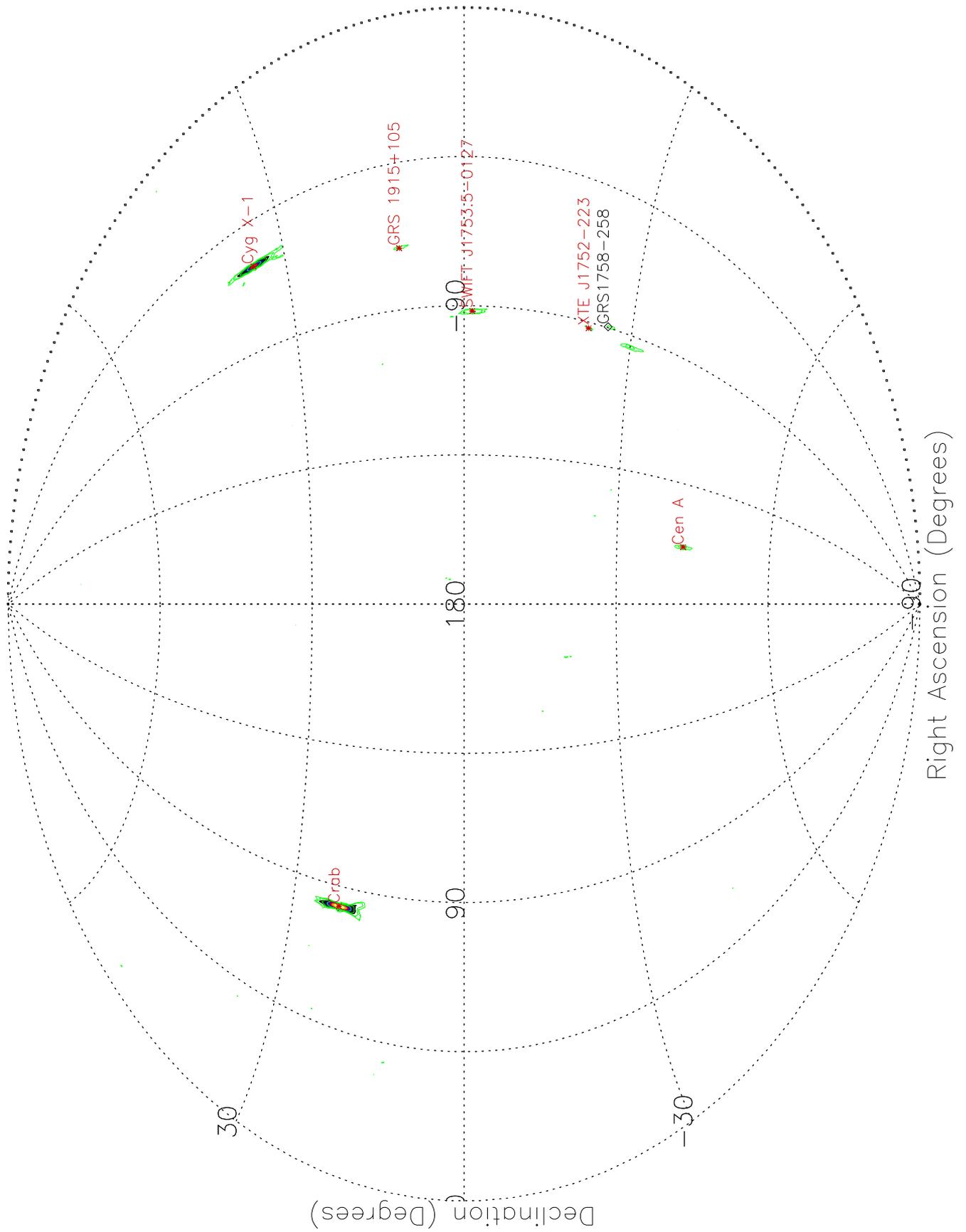}
\vspace{75mm}
\caption{All-sky reconstruction in the 100-300 keV band for 2008 August 12 to 2012 September 14.  Contours have been over plotted at 3. 5,  5,  10,  and 20 sigma.}
\label{fig:map100}
\end{figure*}

\begin{table*}
  \caption{Sources Detected by IDEOM in the 12-50 keV Band}
  \label{table:IDEOM12}
  \centering
  \begin{tabular}{c c c c c c}
   \hline\hline
    Source Name  & Known Position & Measured Position & Position Error & Flux \tablefootmark{1} \\
     & (Degrees) & (Degrees) & (Degrees) & (mCrab)  \\
    \hline
    NGC 1275           & (49.95, 41.52)   & (50.21, 41.25)   & 0.33 &  13.7   \( \pm \) 2.8   \\
    X Per              & (58.85, 31.05)   & (58.92, 31.00)   & 0.08 &  35.6   \( \pm \) 2.8   \\
    LMC X-4            & (83.20, -66.36)  & (83.59, -66.00)  & 0.40 &  24.2   \( \pm \) 2.8   \\
    Crab               & (83.63, 22.01)   & (83.59, 22.00)   & 0.05 & 1000.0  \( \pm \) 2.8   \\
    LMC X-3            & (84.79, -64.08)  & (84.79, -63.75)  & 0.33 &   8.4   \( \pm \) 2.8   \\   
    4U 0614+09         & (94.28, 9.13)    & (94.16, 9.00)    & 0.18 &  29.1   \( \pm \) 2.8   \\
    Vela Pulsar        & (128.85, -45.18) & (128.55, -45.25) & 0.22 &  12.8   \( \pm \) 2.8   \\ 
    Vela X-1           & (135.52, -40.55) & (135.63, -40.75) & 0.21 &  199.6  \( \pm \) 2.8   \\
    NGC 3227           & (155.88, 19.86)  & (155.66, 19.75)  & 0.23 &  7.6    \( \pm \) 2.7   \\
    Cen X-3            & (170.31, -60.62) & (170.59, -60.50) & 0.18 &  112.9  \( \pm \) 2.8   \\
    1E 1145.1-6141     & (176.86, -61.95) & (176.79, -62.00) & 0.06 &  18.3   \( \pm \) 2.9   \\
    NGC 4151           & (182.65, 39.41)  & (182.40, 39.25)  & 0.25 &  25.5   \( \pm \) 2.8   \\
    NGC 4388           & (186.44, 12.66)  & (186.68, 12.50)  & 0.28 &  11.6   \( \pm \) 2.7   \\ 
    GX 301-2           & (186.65, -62.77) & (186.68, -63.00) & 0.23 &  151.4  \( \pm \) 2.9   \\
    3C 273             & (187.27, 2.05)   & (187.34, 1.75)   & 0.31 &  13.9   \( \pm \) 2.8   \\
    Coma Cluster       & (194.95, 27.98)  & (195.09, 28.00)  & 0.12 &  8.1    \( \pm \) 2.7   \\
    GX 304-1           & (195.32, -61.60) & (195.43, -61.50) & 0.11 &  38.9   \( \pm \) 2.9   \\
    NGC 4945           & (196.36, -49.47) & (196.17, -49.75) & 0.31 &  11.7   \( \pm \) 2.8   \\
    Cen A              & (201.36, -43.01) & (201.27, -43.50) & 0.49 &  49.7   \( \pm \) 2.8   \\
    4U 1323-62         & (201.65, -62.14) & (201.82, -62.00) & 0.16 &  13.5   \( \pm \) 2.8   \\
    NGC 5252           & (204.57, 4.54)   & (204.56, 4.56)   & 0.71 &  7.4    \( \pm \) 2.8   \\
    IC 4329 A          & (207.33, -30.31) & (207.22, -30.25) & 0.11 &  9.2    \( \pm \) 2.8   \\
    Circinus Galaxy    & (213.29, -65.34) & (213.18, -65.25) & 0.10 &  14.0   \( \pm \) 2.8   \\
    NGC 5506           & (213.31, -3.21)  & (213.29, -3.00)  & 0.21 &  10.1   \( \pm \) 2.8   \\
    Sco X-1            & (244.98, -15.64) & (244.99, -16.00) & 0.36 &  2688.8 \( \pm \) 3.0   \\
    4U 1626-67         & (248.07, -67.46) & (248.25, -67.50) & 0.08 &  54.8   \( \pm \) 2.9   \\
    Her X-1            & (254.45, 35.34)  & (254.57, 35.50)  & 0.18 &  69.5   \( \pm \) 2.8   \\
    OAO 1657-415       & (255.19, -41.67) & (255.78, -40.50) & 1.25 &  42.1   \( \pm \) 3.1   \\
    4U 1700-377        & (255.98, -37.84) & (256.39, -39.00) & 1.20 &  31.3   \( \pm \) 3.3   \\
    GX 349+2           & (256.45, -36.41) & (256.06, -37.00) & 0.66 &  197.4  \( \pm \) 3.1   \\
    GRS 1724-308       & (261.89, -30.80) & (261.81, -30.75) & 0.09 &  51.5   \( \pm \) 3.8   \\
    GX 9+9             & (262.93, -16.96) & (263.02, -17.75) & 0.79 &  38.9   \( \pm \) 4.3   \\
    GX 1+4             & (263.00, -24.74) & (262.88, -25.00) & 0.28 &  46.2   \( \pm \) 3.5   \\
    GX 354-0           & (263.35, -33.39) & (263.36, -32.75) & 0.64 &  29.9   \( \pm \) 3.4   \\
    X1735-444          & (264.74, -44.45) & (264.41, -43.75) & 0.74 &  17.8   \( \pm \) 3.2   \\
    1E 1740.7-2942     & (265.97, -29.74) & (266.74, -30.00) & 0.72 &  70.4   \( \pm \) 4.8   \\
    MAXI J1745-288     & (266.46, -28.82) & (266.55, -28.50) & 0.33 &  61.0   \( \pm \) 4.8   \\
    IGR J17464-3213    & (266.57, -32.24) & (266.82, -31.50) & 0.77 &  30.1   \( \pm \) 4.7   \\
    SWIFT J1753.5-0127 & (268.36, -1.45)  & (268.71, -2.25)  & 0.87 &  41.9   \( \pm \) 3.6   \\
    GX 5-1             & (270.27, -25.08) & (270.33, -25.00) & 0.10 &  215.9  \( \pm \) 10.7  \\
    GX 9+1             & (270.38, -20.53) & (270.57, -20.00) & 0.56 &  47.1   \( \pm \) 6.4   \\
    GX 13+1            & (273.63, -17.15) & (273.82, -17.25) & 0.20 &  52.0   \( \pm \) 8.8   \\
    4U 1812-12         & (273.80, -12.08) & (273.48, -12.25) & 0.35 &  54.8   \( \pm \) 8.3   \\
    GX 17+2            & (274.00, -14.03) & (273.93, -14.25) & 0.23 &  165.7  \( \pm \) 6.2   \\
    H1820-303          & (275.92, -30.36) & (275.93, -30.50) & 0.13 &  81.3   \( \pm \) 4.1   \\
    3A 1822-371        & (276.44, -37.10) & (276.83, -36.25) & 0.91 &  23.1   \( \pm \) 3.2   \\
    GS 1826-238        & (277.36, -23.79) & (277.29, -23.00) & 0.80 &  46.1   \( \pm \) 4.7   \\
    Ser X-1            & (279.99, 5.03)   & (280.02, 4.25)   & 0.79 &  38.3   \( \pm \) 2.9   \\
    HT 1900.1-2455     & (285.03, -24.92) & (285.06, -25.75) & 0.83 &  18.7   \( \pm \) 2.9   \\
    GRS 1915+105       & (288.82, 10.97)  & (288.81, 11.00)  & 0.04 &  419.4  \( \pm \) 3.1   \\
    Cyg X-1            & (299.59, 35.20)  & (299.70, 35.25)  & 0.10 &  525.7  \( \pm \) 2.8   \\
    EXO 2030+375       & (308.06, 37.63)  & (308.19, 37.50)  & 0.17 &  26.3   \( \pm \) 3.0   \\
    Cyg X-3            & (308.10, 40.95)  & (308.07, 41.00)  & 0.05 &  187.5  \( \pm \) 3.0   \\
    IGR J21247+5058    & (321.18, 50.98)  & (321.38, 51.00)  & 0.13 &  11.7   \( \pm \) 2.8   \\
    Cyg X-2            & (326.17, 38.32)  & (325.98, 38.25)  & 0.16 &  84.3   \( \pm \) 2.8   \\
    4U 2206+54         & (331.99, 54.52)  & (331.92, 54.50)  & 0.04 &  9.6    \( \pm \) 2.8   \\
    \hline
  \end{tabular}

\tablefoot{
  \tablefoottext{1}{1000 mCrab = 0.486 ph/cm\textsuperscript{2}/s}
}
\end{table*}

\renewcommand{\arraystretch}{1.0}
\begin{table*}
  \caption{Sources Detected by IDEOM in the 50-100 keV Band}
  \label{table:IDEOM50}
  \centering
  \begin{tabular}{c c c c c c}
   \hline\hline
    Source Name  & Known Position & Measured Position & Position Error & Flux \tablefootmark{1}  \\
     & (Degrees)  & (Degrees) & (Degrees) & (mCrab)   \\
    \hline
    X Per              & (58.85, 31.05)   & (58.92, 31.00)   & 0.08  & 36.1 \( \pm \) 2.6   \\
    Crab               & (83.63, 22.01)   & (83.59, 22.00)   & 0.05  & 1000.0 \( \pm \) 2.6   \\
    4U 0614+09         & (94.28, 9.13)    & (94.42, 8.00)    & 1.15  & 17.6 \( \pm \) 2.7   \\
    Vela X-1           & (135.52, -40.55) & (135.30, -40.75) & 0.26  & 19.7 \( \pm \) 2.8   \\
    NGC 4151           & (182.65, 39.41)  & (182.40, 39.25)  & 0.25  & 33.4 \( \pm \) 2.7   \\
    NGC 4388           & (186.45, 12.66)  & (185.86, 10.50)  & 2.24  & 14.0 \( \pm \) 2.6   \\
    3C 273             & (187.61, 2.00)   & (187.27, 2.05)   & 0.34  & 19.1 \( \pm \) 2.6    \\
    Cen A              & (201.36, -43.01) & (201.48, -43.25) & 0.25  & 71.2 \( \pm \) 2.8   \\
    GX 339-4           & (255.70, -48.78) & (255.56, -48.75) & 0.10  & 21.9 \( \pm \) 3.3   \\
    4U 1700-377        & (255.98, -37.84) & (255.88, -38.50) & 0.66  & 93.2 \( \pm \) 2.9   \\
    GX 1+4             & (263.00, -24.74) & (262.90, -24.75) & 0.10  & 49.5 \( \pm \) 3.6   \\
    1E 1740.7-2942     & (265.96, -29.74) & (265.96, -30.25) & 0.51  & 106.7 \( \pm \) 5.7   \\
    MAXI J1745-288     & (266.46, -28.82) & (266.49, -28.25) & 0.57  & 38.3 \( \pm \) 4.2   \\
    IGR J17464-3213    & (266.57, -32.24) & (266.07, -31.75) & 0.65  & 34.3 \( \pm \) 5.4   \\
    XTE J1752-223      & (268.04, -22.32) & (268.06, -23.00) & 0.68  & 26.2 \( \pm \) 5.0   \\
    SWIFT J1753.5-0127 & (268.36, -1.45)  & (268.34, -1.50)  & 0.06  & 93.8 \( \pm \) 3.1   \\
    GRS 1758-258       & (270.30, -25.73) & (270.36, -26.00) & 0.27  & 52.8 \( \pm \) 3.9   \\
    4U 1812-12         & (273.80, -12.08) & (273.23, -11.50) & 0.81  & 32.5 \( \pm \) 3.8   \\
    GX 17+2            & (274.00, -14.03) & (274.03, -14.75) & 0.71  & 26.6 \( \pm \) 3.8   \\
    GS 1826-238        & (277.36, -23.79) & (277.49, -24.00) & 0.23  & 52.8 \( \pm \) 2.8   \\
    GRS 1915+105       & (288.82, 10.97)  & (288.82, 10.75)  & 0.22  & 136.9 \( \pm \) 2.8   \\
    Cyg X-1            & (299.59, 35.20)  & (299.70, 35.00)  & 0.22  & 698.6 \( \pm \) 2.9   \\
    \hline
  \end{tabular}
\tablefoot{
  \tablefoottext{1}{1000 mCrab = 0.068 ph/cm\textsuperscript{2}/s}
}
\end{table*}
\begin{table*}
  \caption{Sources Detected by IDEOM in the 100-300 keV Band}
  \label{table:IDEOM100}
  \centering
  \begin{tabular}{c c c c c c}
   \hline\hline
    Source Name  & Known Position & Measured Position & Position Error & Flux \tablefootmark{1} \\
                 & (Degrees) & (Degrees) & (Degrees) & (mCrab)  \\
    \hline
    Crab               & (83.63, 22.01)    & (83.59, 22.00)   & 0.05  & 1000.0 \( \pm \) 5.3  \\
    Cen A              & (201.36, -43.01)  & (201.34, -43.00) & 0.03  & 86.0   \( \pm \) 5.6  \\
    1E 1740.7-2942     & (265.96, -29.74)  & (266.25, -30.25) & 0.57  & 97.1   \( \pm \) 6.1  \\
    SWIFT J1753.5-0127 & (268.36, -1.45)   & (268.34, -1.50)  & 0.06  & 110.5  \( \pm \) 5.6  \\
    GRS 1758-258       & (270.30, -25.73)  & (270.41, -26.50) & 0.78  & 54.9   \( \pm \) 6.1  \\
    GRS 1915+105       & (288.82, 10.97)   & (288.81, 11.00)  & 0.04  & 61.0   \( \pm \) 5.5  \\
    Cyg X-1            & (299.59, 35.20)   & (299.70, 35.00)  & 0.22  & 650.5  \( \pm \) 5.7  \\
    \hline
  \end{tabular}
\tablefoot{
  \tablefoottext{1}{1000 mCrab = 0.035 ph/cm\textsuperscript{2}/s}
}
\end{table*}

Sources detected by IDEOM and verified by EOT in the 12-50 keV band are listed in Table~\ref{table:IDEOM12} with the fluxes as determined by the IDEOM analysis.  In the 12-50 keV band, the average flux for the Crab over 2008 August 12 to 2012 September 14 is 0.486 ph/cm\textsuperscript{2}/s defined for the purpose of this paper to be 1000 mCrab.  The fluxes in the table have been listed for convenience in units of mCrab.  In this energy range, 67 sources were detected in the GBM-EOT daily analysis while IDEOM found 56 sources, including 17 sources found from Method 1 and 42 sources from Method 2 with 3 sources detected by both methods.  Of the 42 sources detected by IDEOM Method 2, all were already included in the EOT catalog.  Two of the sources detected by IDEOM are in the GBM input catalog but are not detected in the GBM-EOT daily analysis.  Of the 26 sources detected by the daily GBM-EOT analysis but not found by IDEOM, 11 were close to a brighter source (within a few degrees) and thus could not be separated, 12 did not show a peak feature above \( 3.5 \sigma\) in the image, three did not meet the detection threshold of \( > 14 \sigma\) for the integrated significance, and one (A0535+262) lacked image data during outburst and was not expected to be detected.  The daily GBM-EOT analysis detected 17 transient sources in this energy range; 9 of these were detected by IDEOM.  

There were 10 sources with an average GBM-EOT flux \( < 25 \) mCrab detected by the daily GBM-EOT analysis.  Only three of those were found with IDEOM.  For the 26 sources with an average flux between 25 mCrab and 50 mCrab, 13 were missed by IDEOM.  Three sources were in the Galactic Center region and blended with other sources, two were blended with GRS 1915+105, five are confused with Sco X-1, two had an integrated significance \(< 14\), and one was A0535+262.  Above 50 mCrab, 31 sources are detected by the daily GBM-EOT analysis.  All but three are also seen by IDEOM.  The IDEOM sensitivity depends on position on the sky. For sources not near bright sources and their arms, though, the distinct difference in the number of sources detected above and below 50 mCrab suggests an IDEOM sensitivity of $\sim 50$ mCrab. For the two sources in the GBM catalog found by IDEOM but not by the daily GBM-EOT analysis, 3C 273 and GRS 1724-308 have daily GBM-EOT significances of \( 4.8 \sigma \text{ and } 4.7 \sigma\), respectively.  The corresponding IDEOM significances are \( 5.0 \sigma \text{ and } 13.5 \sigma\).  The daily GBM-EOT analysis and IDEOM fluxes for 3C 273 are consistent within the errors.  For GRS 1724-308, the IDEOM flux is 3.5 times larger than the daily GBM-EOT analysis flux, likely due to contributions from source interference in the crowded Galactic Center region.

Table~\ref{table:IDEOM50} and Fig.~\ref{fig:map50} show the sources detected by IDEOM in the 50-100 keV band.  For clarity, sources along the Galactic Ridge are again identified by number in Table~\ref{galridge}.  The average flux for the Crab in the 50-100 keV energy range is 0.068 ph/cm\textsuperscript{2}/s.  In this energy range, the daily GBM-EOT analysis detects 25 sources while IDEOM detects 22 sources, three of which are in the GBM catalog but not detected by the daily GBM-EOT analysis.  There were 6 transients detected by the daily GBM-EOT analysis with 5 of them found by IDEOM.  Three of the six sources not detected by IDEOM have no significant feature in the image (NGC 2110, Sco X-1, and GX 354-0).  Of the other three, two sources were close to a detected source and unable to be separated by IDEOM (OAO 1657-415 and 1A 1742-294) and one (Cyg X-3) had an integrated significance \( < 14\).  The average flux of the three undetected sources without an image feature is less than 22 mCrab.  The three  sources detected by IDEOM but not by the daily GBM-EOT analysis were NGC 4388, GRS 1758-258, and 4U 1812-12.  The daily GBM-EOT analysis sensitivity is low for sources close together as the number of usable occultation steps for each source is reduced.  As described in Sec.\ 3.3, IDEOM does not exclude occultation steps for sources close together on the sky when generating an image, allowing GRS 1758-258 (\( \sim 0.75^{\circ} \) away from GX 5-1) and 4U1812-12 (\( \sim 1.9^{\circ} \) away from GX 17+2) to be detected above 50 keV by IDEOM while they are detected only in the 12-50 keV band in the daily GBM-EOT analysis.  \emph{INTEGRAL} observations have shown spectra of GRS 1758-258 and 4U 1812-12 with significant flux out to 100 keV \citep{Pottschmidt2008,Tarana2006} and no significant flux in this range from GX 5-1 and GX 17+2 \citep{Paizis2005,Mainardi2010,Migliari2007},  in agreement with the results from IDEOM.  The third source, NGC 4388, is detected at \( 5.4 \sigma \) by IDEOM and \( 4.7 \sigma \) by the daily GBM-EOT analysis with flux measurements consistent within the errors between the two methods.  

Table~\ref{table:IDEOM100} shows the sources found by IDEOM in the 100-300 keV band.  For the 100-300 keV energy range, the average flux of the Crab is 0.035 ph/cm\textsuperscript{2}/s.  In this energy range, the daily GBM-EOT analysis detected six sources.  No transient sources were detected by  the daily GBM-EOT analysis or IDEOM in this energy band.  IDEOM found all six of these sources and also detected GRS 1758-258.  The source with the lowest average flux is GRS 1758-258 at \( \sim 54 \) mCrab with an IDEOM peak significance of  \( 9.0 \sigma\).  Although GRS 1758-258 is included in the predetermined EOT catalog, the daily GBM-EOT analysis significance for this source is \(1.6 \sigma\) with a flux of 58 mCrab.  The difference in significance is due to source crowding and thus fewer occultation measurements for the daily GBM-EOT analysis.  There are two sources detected by IDEOM between \( 3.5 \sigma \text{ and } 5 \sigma \) that are not detected by the daily GBM-EOT analysis.  These are 3C 273 with an average flux of \( \sim 20 \) mCrab and a significance of \( 3.9 \sigma \) and XTE J1752-223 with an average flux of \( \sim 32 \) mCrab and a significance of \( 4.6 \sigma \).  The lack of unmodeled sources at these energies suggests that the catalog is complete down to at least 50 mCrab in the 100-300 keV band and that the discrepancies between the EOT and EBOP analyses of BATSE occultation data are likely not due to an incomplete catalog but rather to inaccuracies in the EBOP background model.

Tables 3-5 show the source position reconstructed by IDEOM together with the known position of the source. At 12-50 keV, the distribution of source position errors has a half width at half maximum \( \sim 0.2^{\circ}\), consistent with the size of the IDEOM virtual source sky pixels. The distribution then has a tail extending beyond \(1^{\circ}\) corresponding mainly to sources in crowded regions on the sky (e.g., the Galactic Plane) or along arms associated with bright sources.

\section{Sources Detected by GBM and LAT}
Four persistent sources are found by IDEOM that are also detected by the \emph{Fermi}/LAT: NGC 1275, the Crab, 3C 273, and Cen A.  GBM observations can provide long term average flux measurements that are not available with pointed instruments.  The amount of time during a precession period when the \( \beta \) angle is too large, and the source is unable to be occulted, can be estimated \citep{Harmon2002}:
  \begin{equation}
    \tau_{gap} \approx 2 \frac{P_{Precession}}{2 \pi} \cos^{-1} \left(\frac{\sin \theta_{Occ} - \cos i \sin | \delta |}{ \sin i \cos \delta }\right)  .
    \end{equation}
Here \(P_{Precession}\) is the orbital precession period of the satellite, \( \theta_{Occ} \) is the \( \beta \) angle at which occultations are no longer possible, \( i \) is the orbital inclination of the satellite, and \( \delta\) is the declination of the source.  For \emph{Fermi}, \(P_{Precession} \approx 53\) days, \( \theta_{Occ} = 66^{\circ} \), and \( i = 25.6^{\circ} \).  The Crab (\( \delta =22.01^{\circ}\)) and 3C 273 (\( \delta =2.05^{\circ}\)) have occultations throughout a precession period while NGC 1275 (\( \delta =41.51^{\circ} \)) is not occulted for \( \sim 2\) days and Cen A (\( \delta = -43.02^{\circ}\)) is not occulted for \( \sim 5 \) days.  Thus all four sources have flux measurements for a significant part of a precession period.

    Spectra are shown in Fig.~\ref{spectra} for these four sources using 4 years of broad band GBM data, LAT 2FGL catalog data \citep{Nolan2012}, \emph{INTEGRAL} data \citep{Eckert2009,JR2009}, and data from COMPTEL \citep{COMPTEL2000}.  The GBM data span 2008 August 12 to 2012 August 11, LAT data cover 2008 August 4 to 2010 August 1, and COMPTEL data are from 1991 May 16 to 1996 October 15.  \emph{INTEGRAL} observations for NGC 1275 are from 2003 March to 2004 August while observations of the Crab span \( \sim 5.5 \) years from 2003 February to 2008 September.

\begin{figure*}[h!]
\centering
\includegraphics[angle=0,scale=0.825, trim=0mm 5mm 0mm 5mm]{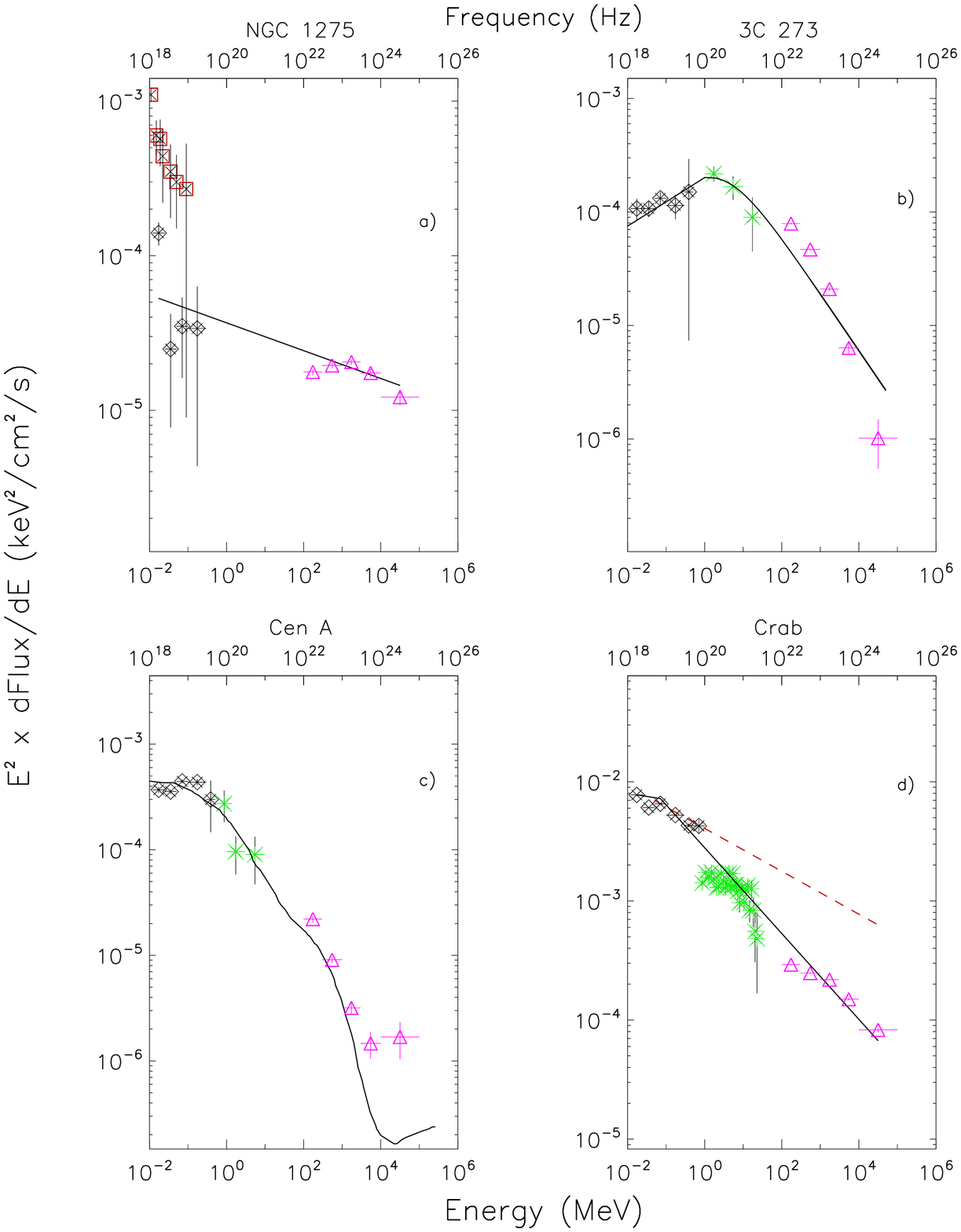}
\vspace{10mm}
\caption{\label{spectra} \emph{INTEGRAL} (red squares), GBM (black diamonds), COMPTEL (green asterisks) LAT (magenta triangles) (a) NGC 1275 spectrum with \emph{INTEGRAL}, GBM, and, LAT data and the \citet{Brown2011} power law spectrum extrapolated from LAT energies down to 12 keV.  (b) 3C 273 spectrum with model of the form presented by \citet{vonMontigny1997} overplotted.  (c) Cen A spectrum with lepto-hadronic model from \citet{Reynoso2011} shown.  (d)Crab spectrum with GBM 3 year average spectrum \citep{WilsonHodge2012} overplotted in (black) solid line, and \emph{INTEGRAL}/SPI spectrum \citep{JR2009} in (red) dashed line.}
\end{figure*}

\subsection{NGC 1275}

NGC 1275 is an elliptical galaxy in the middle of the Perseus galaxy cluster.  HEAO 1 observations showed emission at energies from 10-93 keV with no obvious variability \citep{Primini1981}.  Later observations with \emph{CGRO}/OSSE failed to detect NGC 1275 in the 50-500 keV energy range with upper limits significantly below the fluxes observed during earlier missions \citep{Osako1994}, indicating long-term variability.  \emph{INTEGRAL} has detected the source at 3-20 keV with JEM-X and at 20-120 keV with IBIS/ISGRI \citep{Eckert2009}.  The flux observed by \emph{INTEGRAL} was time variable but lower by a factor of 3 than the extrapolation of the \( E^{-2} \) power law component reported by \emph{Chandra} \citep{Sanders2005}.  \citet{Eckert2009} fit the \emph{INTEGRAL} data from 2003 March to 2004 August with a model of a two-temperature plasma with a central temperature \( kT= 3\) keV plus an AGN contribution.  At higher energies, \emph{CGRO}/EGRET observations resulted in only upper limits \citep{Reimer2003} while the \emph{Fermi}/LAT detects NGC 1275 with variations on monthly timescales with an average best fit power law spectral index of \( \Gamma = 2.13\) \citep{Kataoka2010}.  \citet{Kataoka2010} have also fit the data using a cut-off power law model resulting in a better \( \chi^{2} \) with \(\Gamma = 2.07 \text{ and } E_{C} = 42.2 \) GeV.  \citet{Brown2011} found \( \sim 2\) years of LAT data to be best fit by a power law with \( \Gamma = 2.09 \).  Fig.~\ref{spectra}(a) shows the average LAT power law spectrum from \citet{Brown2011} extrapolated down to GBM energies.  The LAT power-law spectrum extended down to 12-25 keV fits the contemporaneous GBM data reasonably well while the \emph{INTEGRAL} result is roughly an order of magnitude above the GBM data.  The GBM data do not show any significant variability during the mission so far, based on the light curves published on the GBM Occultation Project website\footnote{http://heastro.phys.lsu.edu/gbm/}.  

\subsection{3C 273}
The bright close flat spectrum radio quasar 3C 273 has been a frequent target of X-ray and \( \gamma\)-ray observations since the 1970s.  Observations with the HEAO A2 experiment in the 2-60 keV energy range are best fit with a power law of \( \Gamma =1.41\) \citep{Worrall1979} while HEAO A4 measurements in the 13-180 keV energy range give a slope of \( \Gamma =1.67\) \citep{Primini1979}.  OSSE observations are well fit by a power law model with \( \Gamma = 1.71 \) and do not show significant spectral variability despite flux variability during observations.  With \emph{INTEGRAL}, \citet{Cournoiser2003} fit the 25-100 keV IBIS/ISGRI spectrum to a power law with \( \Gamma =1.95 \pm 0.2\) and the 20-200 keV SPI spectrum to a power law with \( \Gamma = 1.66 \pm 0.28\).  Later \emph{INTEGRAL} and \emph{XMM-Newton} observations \citep{Chernyakova2007} confirmed a softening of the power law spectrum with time, reporting a photon index \( \Gamma = 1.82 \pm 0.01\) with observations from 2003-2005.  Such a soft spectrum can be produced by optically thin inverse Compton emission from relativistic electrons, but the harder spectrum observed by early experiments cannot be produced this way unless, as suggested by \citet{Chernyakova2007}, there is a significant density of protons in the \( \gamma\)-ray emission region.  COMPTEL and EGRET reported a break at \( \sim 1 \) MeV with a slope of \( \Gamma = 2.4 \) above 1 MeV \citep{Johnson1995}.  COS B observations at \( \gamma\)-ray energies (50-800 MeV) show a power law of \( \Gamma = 2.5\) \citep{Bignami1981}.  However, spectral hardening was observed at a time when the source brightened \citep{vonMontigny1997,Collmar2000}.  \emph{Fermi}/LAT has observed variations in the photon index from \( \Gamma = 2.4 \pm 0.2 \text{ to } 3.3 \pm 0.3\) and a clear hardening of the spectrum at times when the flux increases \citep{Soldi2009}.  The GBM light curves show distinct variability over the energy range 12-300 keV.  The emission has been modeled by synchrotron self Compton, inverse Compton, and a proton-initiated cascade model, and \citet{vonMontigny1997} has fit the high-energy spectrum satisfactorily with an empirical model of the form
\begin{equation}
\frac{dN}{dE} = \left( \frac{N}{E_B} \right) \frac{(E/E_{B})^{-(1+ \alpha) }}{1 + (E/E_{B})^{ \beta}}.
\end{equation}
Fig.~\ref{spectra}(b) shows the 4-year average GBM spectrum together with the LAT and COMPTEL data and the \citet{vonMontigny1997} model with \( \alpha = 0.7, \beta = 0.8, \text{ and } E_B = 2.36 \) MeV, corresponding to a steepening from \( \alpha +1 = 1.7 \text{ to } \alpha + \beta +1 = 2.5\).  The model and data agree well up to \( \sim 10 \) GeV.

\subsection{Cen A}
Cen A is one of the brightest AGN detected from radio energies up to the TeV range \citep{Aharonian2009} and in ultra-high energy cosmic rays by Auger \citep{Abraham2009}.  Gamma-rays are produced by the jet in the central core and by inverse Compton scattering of microwave and infrared-optical photons in the giant radio lobes.  At hard X-ray energies,  Ginga and balloon-flight observations revealed a power law fit with a slope of \( \sim 1.8\) with a possible break at \( \sim 180 \) keV \citep{Miyazaki1996}.  OSSE, COMPTEL, and EGRET observations during an intermediate emission state showed a three-segment broken power law with breaks at \(E_{1} = 150\) keV and \(E_{2} = 16.7\) MeV and slopes of \( \Gamma_{1} =1.74, \Gamma_{2} =2.3 \text{, and }\Gamma_{3} =3.3 \), while a fit during a low emission state showed breaks at \(E_{1} = 140\) keV and \(E_{2} = 590\) keV and slopes of \( \Gamma_{1} =1.73, \Gamma_{2} =2.0 \text{ and, }\Gamma_{3} =2.6 \) \citep{Steinle1998}.  BATSE Earth occultation data were fit with a slope of \(\Gamma = 1.84\) and no evidence of a break out to 1 MeV \citep{Wheaton1996}.  INTEGRAL observations by \citet{Beckmann2011} are well fit to an absorbed cut-off power-law model with \( \Gamma = 1.73\) and \( E_{C} = 434\) keV.  The third EGRET catalog reports a power-law fit with \( \Gamma = 2.58\) \citep{Hartman1999}.  LAT measurements show a power-law fit with \( \Gamma = 2.67 \) \citep{Abdo2010}, consistent with previous measurements at these energies.  The observations from LAT energies through the radio can be fit with a synchrotron/synchrotron self-Compton model with a single emission region \citep{Abdo2010} or by a combined lepto-hadronic model in which the hard X-ray/soft \(\gamma\)-ray emission is produced by a combination of electron synchrotron radiation, proton synchrotron, and inverse Compton emission \citep{Reynoso2011}.  In  Fig.~\ref{spectra}(c) the Reynoso et al. lepto-hadronic model fits well with the LAT, COMPTEL, and GBM data over the range from 12 keV to 10 GeV.    

\subsection{Crab}
The Crab has long been considered to be a steady ``standard candle''.  Variability was seen with several instruments on board \emph{CGRO}: COMPTEL and EGRET showed variations in the 1 MeV to 150 MeV range \citep{DeJager1996} and BATSE \citep{LW2003} suggested flux and spectral variability in the 35-300 keV energy range on weekly timescales, though the BATSE short-term variability results were not confirmed.  GBM, \emph{INTEGRAL}, \emph{RXTE}, and \emph{Swift} data have shown that the 15-100 keV flux varies by several percent per year on a timescale of \( \sim 3\) years \citep{WilsonHodge2011}.  GBM has presented an average spectrum of the Crab pulsar plus nebula best fit by a broken power law model with \( \Gamma_1 = 2.06 \pm 0.01 \), \( \Gamma_2 = 2.36 \pm 0.05 \) and a break energy at \( 98 \pm 9\) keV \citep{WilsonHodge2012}. \emph{INTEGRAL} \citep{JR2009} fit \( \sim 5.5 \) years of SPI data to a broken power law spectrum covering 20 keV to 6 MeV with slopes \( \Gamma_{1} = 2.07 \text{ and } \Gamma_{2} = 2.23 \text{ and a break energy fixed at } 100 \) keV.  When the break energy is left as a free parameter,   \( \Gamma_{1} = 2.04 \text{ and } \Gamma_{2} = 2.18 \text{ with a break energy at } 62 \) keV.  Fig.~\ref{spectra}(d) shows the \emph{INTEGRAL} (red dashed) and GBM (solid black) models extended to 10 GeV together with results from COMPTEL and \emph{Fermi}/LAT.  The GBM broken power law spectrum provides excellent agreement up to above 10 GeV.  

\section{Conclusions and Future Work}
All-sky reconstructions of the hard X-ray/soft \(\gamma\)-ray sky have been generated with IDEOM using GBM CTIME data.  These are the first all-sky reconstructions using Earth occultation since \citet{Shaw2004}.  The maps cover the 12-50 keV, 50-100 keV, and 100-300 keV energy bands and span \( \sim 1500 \) days.  43 known sources have been identified by the IDEOM imaging.  Also 17 sources have been added to the GBM EOT catalog through cross-correlating features in maps with sources in the \emph{Swift}/BAT and \emph{INTEGRAL} catalogs.  Most of the calculated source positions from IDEOM are within \( \sim 0.5^{\circ} \) of the known source position, which is the minimum angular resolution as constrained by the Earth's atmosphere.  No source was detected at \( > 5 \sigma\) in the 100-300 keV reconstruction that was absent from the GBM-EOT catalog down to \( \sim 50 \) mCrab.  This result suggests that the discrepancy between the MSFC and JPL BATSE occultation analyses \citep{Harmon2004} is not due to unmodeled sources, but rather that a more likely cause of the difference is  the background model.  Future work with IDEOM includes updating the current images with more recent GBM data to increase sensitivity and continuing to search for additional sources.  

Spectral analysis of the four persistent sources significantly detected by both GBM and LAT (NGC 1275, the Crab, 3C 273, and Cen A) has been performed using GBM, \emph{INTEGRAL}, COMPTEL (where available), and LAT data.  For NGC 1275, the extrapolation of the average \emph{Fermi}/LAT spectrum down to GBM energies shows results consistent with GBM.  For 3C 273 and Cen A, the GBM data are again consistent with model fits \citep{vonMontigny1997,Reynoso2011} extended down from LAT energies.  For the Crab, the fit suggested in \citet{WilsonHodge2012} has been extrapolated up to LAT energies and again provides a good fit.

\begin{acknowledgements}
This work is supported by the NASA Fermi Guest Investigator program, NASA/Louisiana Board of Regents Cooperative Agreement NNX07AT62A (LSU), and the Louisiana Board of Regents Graduate Fellowship Program (J.R.).  This material is based upon work supported by HPC@LSU computing resources as well as the Louisiana Optical Network Institute (LONI).  ACA thanks the support from grants AYA2012-39303, SGR2009-811,  and iLINK2011-0303.

\end{acknowledgements}

\bibliographystyle{/usr/share/texmf/tex/latex/aa/bibtex/aa} 
\bibliography{imaging_paper_Dec12_arxiv_eps} 

\end{document}